\documentclass{appolb}
\usepackage{epsfig}
\usepackage{graphicx}   
\usepackage{rotating}
\usepackage{dcolumn}
\usepackage{bm}
\usepackage{amssymb}
\usepackage{multirow}
\newcommand{\pt}{\mbox{$p_\mathrm{T}\,$}}

\begin{document}
\title{Probe the QGP via dihadron correlations: Jet quenching and Medium-response%
\thanks{Presented at International Symposium On Multiparticle Dynamics (ISMD07)}%
}
\author{Jiangyong Jia
\address{Chemistry Department, Stony Brook University, Stony Brook, NY 11794, USA}
} \maketitle

\begin{abstract}
We summarize the di-hadron correlation results from RHIC, focusing
on the high $p_T$ region and lower $p_T$ region for the away-side.
The former is consistent with fragmentation of jets that surviving
the medium, while the latter suggests the redistribution of the
energy from the quenched jets. We also discuss the role of the jet
in the intermediate $p_T$.
\end{abstract}


\section{High $p_T$ correlation: Jet quenching}
Early results from RHIC has established the creation of the
strongly interaction partonic matter in central Au+Au collisions.
The current focus of the community is to characterize its dynamical
properties. One of the most useful tools at our disposal is the
hard-scattered jets and di-jets. These (di-)jets are created early
in the collision and can subsequently probe the space-time
evolution the QGP via the jet-medium interactions. Both the single-
and away-side di-hadron yield at high $p_T$ are strongly
suppressed, consistent with jet quenching picture, where jets
traversing the medium suffer significant energy loss or even
totally absorbed by the medium.

The strong suppression implies that the center of the medium is
extremely opaque. This and the steeply falling parton spectra
effectively lead to a bias on the energy loss, where the observed
high $p_T$ single hadrons and dihadron pairs mainly come from those
(di)jets that suffer minimal interaction with the medium. In fact,
dihadron data at $p_T>5$ GeV/$c$~\cite{Adams:2006yt,Adare:2007vu}
reveal characteristic jet-like peaks for the near-side
($\Delta\phi\sim0$) and the away-side ($\Delta\phi\sim\pi$),
consistent with fragmentation of almost unmodified jets.

Because the observed jets suffers small energy loss, their
usefulness as a tomography tool is limited. The information about
jet-medium interaction are mostly derived indirectly from the
suppression factors such as $R_{AA}$ and $I_{AA}$. These factors
basically represent the survival probabilities, and they are not
very sensitive to the details of energy loss mechanisms in the high
$p_T$ region). The situation, however, can be somewhat improved if
we combine both single- and dihadron measurements and consider
$R_{AA}$ and $I_{AA}$ simultaneously.

To first order, the suppression of the single hadron yield can be
described by either a downward shift
(absorption)~\cite{Drees:2003zh} or left shift (energy loss with a
fixed fraction)~\cite{Adcox:2004mh}. A more general form can
contain both contributions,
\[
P(\Delta E) = a[(1-b)\delta(\Delta E)+b\delta(\Delta E-E)] +(1-a)\delta(\Delta E-E_0)
\]
If we only consider the downward shift, a simple calculation show
that $I_{AA}>R_{AA}$, simply because there is more matter to
transverse for away-side jet of the survived jets than for single
jets. Since experimental data suggests that $I_{AA}\approx R_{AA}$,
clearly downward-shift only is not sufficient.

On the other hand, the suppression caused by left-shift term
depends on both the energy loss itself as well as on the input
parton spectra shape. The expected binary scaled p+p single hadron
yield (in $dN/dp_T$) has a power-law shape with a power of
8~\cite{Adler:2003pb}; the away side spectra associated with the
leading hadron, however, is much flatter with a power of
4.8~\cite{Jia:2007qi}. For the same amount of left shift, we expect
$I_{AA}<R_{AA}$.

Following the prescription in Ref~\cite{Jia:2007qi}, the fractional
energy loss $S_{loss}$ is related to the suppression factor as
$S_{\rm{loss}} = 1 - (\rm{R_{AA}\; or\; I_{AA}})^{1/(n-1)}$. So if
we require $I_{\rm{AA}}= R_{\rm{AA}}=0.2$, then the away hadron
energy loss fraction would be $S^{I}_{\rm{loss}} = 1 -
0.2^{1/(4.8-1)} = 0.345$, much bigger than the single hadron energy
loss fraction $S^{R}_{\rm{loss}}=1 - 0.2^{1/(8-1)}=0.23$, as
expected (about 50\% more energy loss).

\section{Intermediate and low $p_T$ correlations: Medium response}
Simple energy conservation arguments suggest that the energy of the
quenched jets should be transported to intermediate and low $p_T$
single hadrons or hadron pairs. The shape and yield of such
``jet-induced'' contribution could be sensitive to the properties
of the medium, such as the opacity, transport coefficient, speed of
sound, etc.

\begin{figure}[h]
\begin{tabular}{lr}
\begin{minipage}{0.5\linewidth}
\begin{flushleft}
\epsfig{file=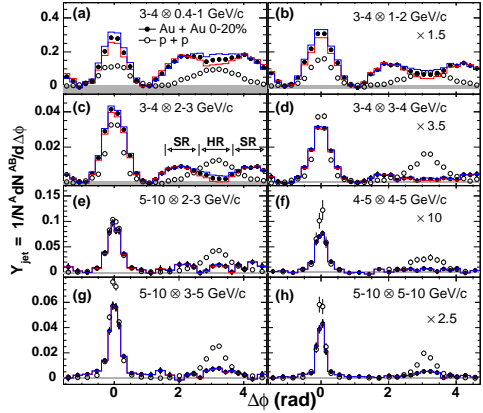,width=1\linewidth}
\end{flushleft}
\end{minipage}
&
\begin{minipage}{0.4\linewidth}
\begin{flushright}
\caption{\label{fig:shape} Per-trigger yield vs. $\Delta\phi$ for
various trigger and partner $p_{T}$ ($p_T^{A}\otimes p_{T}^{B}$),
arranged by increasing pair momentum (sum of $p_T^{A}$ and
$p_{T}^{B}$), in $p+p$ and 0-20\% Au+Au collisions.}
\end{flushright}
\end{minipage}
\end{tabular}
\end{figure}

Many studies show strong modifications of the near- and the
away-side $\Delta\phi$ distributions in this $p_T$ region.
Fig.~\ref{fig:shape} shows a representative subset of the
per-trigger yield distributions arranged by increasing pair
transverse momentum from~\cite{Adare:2007vu}. At low $p_T$, the
near-side jet-induced pairs peak at $\Delta\phi\sim0$, but the peak
is broadened and enhanced with respect to $p+p$ collisions; The
away-side jet-induced pairs peaks at
$\Delta\phi\sim\pi\pm1.1$~\cite{Adare:2007vu,Adler:2005ee,} with a
local minimum at $\Delta\phi\sim\pi$. Going from low to high $p_T$,
the away-side evolves from a broad, roughly flat away-side peak to
a local minimum at $\Delta\phi\sim\pi$. At high $p_T$, jet shape
for Au+Au gradually becomes peaked as for $p+p$, albeit suppressed.
These modification patterns reflect characteristics of the energy
transport of the quenched partons in both \pt and $\Delta\phi$.

The evolution pattern of the away-side jet shape with \pt suggest
separate contributions from a medium-induced component centered at
$\Delta\phi \sim \pi\pm1.1$ (shoulder region, SR) and a
fragmentation component centered at $\Delta\phi \sim \pi$ (head
region, HR). A detailed analysis of this $p_T$ dependence of the
away-side modification in the two regions can be found
in~\cite{Adare:2007vu}. The enhancement of the SR seems to be
limited at $\pt<4$ GeV/$c$, and its yield shows a universal slope
independent of trigger \pt. The suppression of the HR yield
reflects the jet energy loss.

The patterns of the near-side jet shape and yield in
Fig.\ref{fig:shape} also suggest enhancement and broadening at
intermediate \pt. This medium modification was shown to be related
to a long range correlation component in
$\Delta\eta$~\cite{Adams:2004pa}. It was shown to be flat up to
$|\Delta\eta|\sim2$ and was refered to as the $\eta$ ``Ridge''. The
ridge yield seems to dominate over the jet yield at $\pt<4$
GeV/$c$, and it has a universal spectra slope independent of
trigger $p_T$ as well. All these observations suggest the shoulder
yield and near-side ridge yield reflects some intrinsic properties
of the medium.

\begin{figure}[h]
\begin{tabular}{lr}
\begin{minipage}{0.6\linewidth}
\begin{flushleft}
\epsfig{file=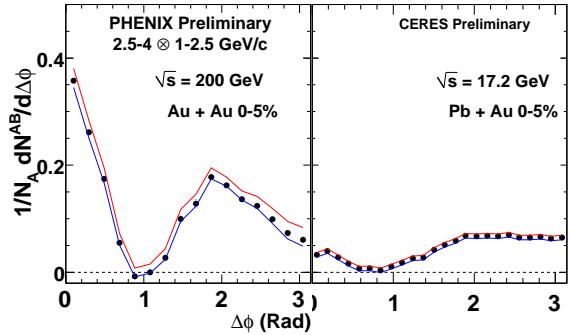,width=1\linewidth}
\end{flushleft}
\end{minipage}
&
\begin{minipage}{0.35\linewidth}
\begin{flushright}
\caption{\label{fig:jetsqrt} a) Per-trigger yield in 0-5\% central
Au+Au collisions from PHENIX at $\sqrt{s_{\rm{NN}}} =$ 200 GeV. b)Per-trigger yield in 0-5\% Pb+Au collisions at $\sqrt{s_{\rm{NN}}}
=$ 17.3 GeV from CERES~\cite{Ploskon:2007es}.}
\end{flushright}
\end{minipage}
\end{tabular}
\end{figure}

Further insight into the physics that drives the SR yield can be
obtained by the studying its $\sqrt{s}$ dependence. In particular,
it is interesting to see whether the two-component picture applies
at much lower collision energy. Results from the two collision
energies ($\sqrt{s_{\rm NN}}$ = 200 and 17.2 GeV from
PHENIX~\cite{Jia:2007tu} and CERES~\cite{Ploskon:2007es}) for
$1<p_{T}^B<2.5<p_T^A<4$ GeV/$c$ are shown side by side in
Fig.~\ref{fig:jetsqrt}. The away-side shapes are strongly
non-Gaussian in both cases. But the 17.2 GeV data looks almost flat
at the away-side.

The 17.2 GeV data is from CERES and is carried out in
$0.1<\eta<0.7$ in the CM frame at $\sqrt{s_{NN}} = 17.3$
GeV~\cite{Ploskon:2007es}. Its pseudo-rapidity widow of 0.6 is
close to 0.7 for PHENIX, thus the corresponding jet yield can be
compared with PHENIX results after applying an upward correction of
0.7/0.6 = 1.17. The maximum of SR in CERES is about half that of
the PHENIX value, whereas the yield at the HR is surprisingly close
to the PHENIX value. The former might suggest a weaker medium
effect at lower energy, while the latter could be a combined result
of a lower jet multiplicity and a weaker jet quenching at SPS.
Further detailed study of the energy dependence of the
punch-through and the medium-induced components can elucidate the
onset of jet quenching and medium response.

Many mechanisms for this energy transport have been proposed for
the near-side~\cite{Voloshin:2004th,Chiu:2005ad,Armesto:2004pt,
Romatschke:2006bb,Majumder:2006wi,Shuryak:2007fu,Wong:2007pz,Pantuev:2007sh}
 and away-side~\cite{Armesto:2004pt,Romatschke:2006bb, Majumder:2006wi,Armesto:2004vz,
Chiu:2006pu, Vitev:2005yg,Polosa:2006hb,Dremin:1979yg,
Koch:2005sx,Stoecker:2004qu,Casalderrey-Solana:2004qm}. Most of
these models are quite qualitative in nature. They typically focus
on either jet shape or jet yield, near-side or away-side, high
$p_T$ or low $p_T$. The fact that both near- and away-side
distributions are enhanced and broadened at low $p_T$ and that the
modifications limited to $p_T\lesssim4$ GeV/$c$, above which the
jet characteristics qualitatively approach jet fragmentation, may
suggest that the modifications mechanisms for the near- and
away-side are related. A model framework including both jet
quenching and medium response, which can describe the full $p_{\rm
T}$ evolution of the jet shape and yield at both near- and
away-side is required to understand the parton-medium interactions.
Our data provide valuable guidance for such future model
development.

\end{document}